\documentclass[referee]{aa} % for a referee version
\usepackage{graphicx}
\usepackage{txfonts}
\begin{document}
   \title{Time--resolved {\it FUSE} Photometric and Spectroscopic Observations: PG\,1219$+$534, PG\,1605$+$072, and PG\,1613$+$426.}

   \subtitle{}

   \author{ Kuassivi
          \inst{1}
          \and
          A. Bonanno
	\inst{2}
	\and
	R. Ferlet
	\inst{3}
          }

   \offprints{Kuassivi}

   \institute{AZimov association\\
	        14, rue Roger Moutte, 83270 St Cyr, France.\\
              \email{martial.andre@free.fr}
         \and 
             INAF--Osservatorio Astrofisico di Catania\\
             Citta Universitaria, 5123 Catania, Italy\\
             \email{abo@ct.astro.it}
	   \and
             Institut d'Astrophysique de Paris\\
	       98 bis, bvd Arago, 75014 Paris, France.\\
             \email{ferlet@iap.fr}
          }

   \date{Received November 15, 2004; accepted March 16, 2005}

   \abstract{
	We report on the detection of pulsations in the far ultraviolet (hereafter FUV) light curves of PG\,1219$+$534, PG\,1605$+$072, and PG\,1613$+$426
obtained with the Far Ultraviolet Spectroscopic Explorer ({\it FUSE}) in time--tagged mode (TTAG). Exposures of the order of
a few Ksecs were sufficient to observe the main frequencies of PG\,1219$+$534 and PG\,1605$+$072 and confirm the detection
of a pulsation mode at the surface of PG\,1613$+$426 as reported from ground. For the first time we derive time--resolved
spectroscopic {\it FUSE} data of a sdB pulsator (PG\,1605$+$072) and comment its line profile variation diagram (hereafter lpv diagram).
We observe the phase shift between the maximum luminosity and the maximum radius to be consistent with the model of an adiabatic pulsator.
We also present evidence that the line broadening previously reported is not caused by rotation but is rather an observational bias due to the
rapid Doppler shift of the lines with 17 km.s$^{-1}$ amplitude. Thus our observations do not support the previous claim that PG\,1605$+$072
is (or will evolve into) an unusually fast rotating degenerate dwarf. These results demonstrate the
asteroseismological potential of the {\it FUSE} satellite which should be viewed as another powerful means of
investigation of stellar pulsations along with the {\it {\it MOST}} and {\it {\it COROT}} missions.

   }

   \maketitle
%
%________________________________________________________________

\section{Introduction}

   Subluminous B(sdB) stars dominate the population of faint blue stars of our own Galaxy and are numerous enough to
account for the UV upturn phenomenon (Brown et al., 2000) observed  in elliptical galaxies and galaxy bulges. Since the discovery
of sdB stars in the globular cluster NGC\,67524 (Heber et al., 1986), evidence has accumulated that sdB stars represent late stage of
stellar evolution. These are evolved objects with typical helium (He) burning cores of 0.5 M$_{\odot}$ surrounded by a thin
H surface layer (less than 2 \% of the mass) and are located near the extreme horizontal branch (EHB) with effective
surface temperatures ranging from 20,000 K to 40,000 K. Although important questions remain regarding the exact evolutionary
path and time-scales, sdB stars are widely believed to be immediate progenitors of low mass white dwarfs. Most of
these subdwarfs reside in close binaries which suggests a formation scenario involving binary interaction and leads
to the general expectation that these associations are progenitors of type Ia supernovae.

Since 1997, the discovery of multi-mode, short period (P = 2--10 min) oscillations among sdBs (Kilkenny et al., 1997)
has provided an unique opportunity for probing their interiors using asteroseismological methods. From a theoretical ground
(Charpinet et al, 1996), the sdB
instability strip has long been predicted to occur between 29,000K and 37,000 K which seems in good agreement with recent
observations. However, relatively few sdB stars in this temperature range are reported to show luminosity variations
and it is not clear whether this is a bias due to the poor detection limit from the ground or the effect of some intrinsic
physical process. The limitation set by the atmospheric scintillation makes mandatory the use of space based observatories
for further asteroseismological investigations.

Many high--resolution FUV spectra of sdB stars have been made available thanks to the {\it FUSE} satellite (Fontaine \& Chayer, 2004).
But it is only recently that detections of pulsation modes with the {\it FUSE} detector have been reported by Kuassivi et al. (2004)
towards PG\,1219$+$534. Several previous successfull detections of stellar pulsation from space with the {\it Hubble Space Telescope}
({\it HST}) have been also reported in the literature (Kepler et al., 2000) with similar performances. However, until now the lack
of high--speed time--resolved spectroscopic data prevented any direct evidences for mode identification from being collected.

The advent of the {\it COROT} space mission (Baglin et al., 2001) will
improve by many orders of magnitude the sensitivity of photometric investigations and bring important clues in the
general field of stellar pulsations. However, we shall point out that even the most awaited {\it COROT} space mission, which is designed as a high
precision, wide field, stellar photometric instrument, will not provide time--resolved spectroscopic data
and will only focus on stellar pulsators brighter than the 9$^{\rm th}$ magnitude thus excluding many faint
oscillators and most (if not all) sdB stars from further investigation.

In this paper we make use of the unexploited {\it FUSE} photometric possibilities and provide an analysis of FUV photometric
data collected towards three sdB stars. The observation and the special reduction technique are described in \S 2 and \S 3. The frequency
analyzes are summarized in \S 4 for each target. In \S 5, we focus on the bright PG\,1605$+$072 and derive
the first lpv diagram ever obtained for a sdB star. Future prospects for space--based asteroseismology with the {\it FUSE} satellite
are discussed in \S 6 and a summary is given in \S 7.

%__________________________________________________________________

\section{Observation}

The {\it FUSE} design consists of four co--aligned optical channels (LiF1--2 and SiC1--2), each channel is made up of
a telescope primary mirror (35 cm aperture), a Focal Plane Assembly (FPA) containing the spectrograph entrance
apertures, a holographic diffraction grating, and a portion of a detector. The {\it FUSE} mission, its planning,
and its on--orbit performance are discussed by Moos et al. (2000) and Sahnow et al. (2000).

A summary of the {\it FUSE} observation used in this work is shown in Table. 1.
The integrated spectra computed by the {\it FUSE} pipeline (CalFUSE 2.4) allow for a spectral resolution of 20,000
and a signal--to--noise per resolution element (0.06 \AA) varying from 5 for PG\,1219$+$534 to 10 for PG\,1605$+$072.

\section{Data reduction}

The reduction of the raw two--dimensional {\it FUSE} detector images into useful FUV photometric series has already
been described in Kuassivi et al. (2004). It proceeds in the three following steps.

First, the photometric time--series is flagged for known instrumental defects correlated with the satellite orbital
motion such as pointing drifts and burst events. Pointing drifts are caused by thermal fluctuations within the
satellite that  perturb the fine co--alignment of the four optical channels. This phenomenon infrequently results
in the loss of one or two channels but rarely affect observations made through the large aperture (LWRS).
Intermittent increases in the count rate, known as burst events, are a more important issue for photometric calibration and need
to be cautiously screened. These events are due to scattered light entering the telescope and occur preferentially
in orbital morning. Since the burst events fully illuminate the detector, a few off spectrum regions
are carefully monitored so as to characterize these burst events in amplitude and duration. The same regions serve
also to monitor the background arrays that are found to be essentially constant throughout the exposures. 

The second step consists in flagging regions of the raw image affected by detector oddities such as dead--pixels or moving shadows. At variance
with dead pixels whose positions are cataloged and whose impact on the total observed flux is negligible, the moving
shadows, known as the ``worm effects'', can reduce the flux by as much as 50 \% over a narrow spectral
range (typically between 110 and 120 nm). Such structures are observed in the LiF 1A, LiF 1B and LiF 2A channels for
all our observations and result in a net loss of about 10\% of the total flux.

The last step consists in excluding the airglow emission features whose orbital modulation affect the
signal.  The photometric time--series are then computed from the photon count rate within these
cleaned regions of the detector.

The same procedure is performed for each detector and each exposure. Although most of the reduction needs to be
done by hand, we would like to point out that many user--friendly IDL applets are available to handle FUV raw data
at the {\it FUSE} web site. In particular, the {\it FUSE\_SCAN} procedure allows to go through each of the above--mentioned
steps.

\section{Frequency analysis}

\subsection{PG\,1219$+$534}
{\it FUSE} time--tagged data have been obtained for the known short period pulsating sdB type star PG\,1219$+$534 (Koen et al., 1999). The
data were acquired on the 16$^{\rm th}$ January 2001 for 6,300 seconds and were reduced following the special procedure already
outlined by Kuassivi et al. (2004) (see section 3). Fig. 1 shows the Direct Fourier Transform (DFT) periodogram computed for a continuous exposure
of 3,200 s and the confidence level for the main frequency.
A simultaneous non--linear least square fit of two sinusoids provide the two main frequencies: 6.9 mHz and 7.8 mHz.
Our result is consistent, within our frequency resolution (0.2 mHz), with the result from ground observations (Koen et al., 1999).
Hence, the two peaks observed in the {\it FUSE} periodogram (Fig. 1) are indeed pair of unresolved components: 6.7--7.0 mHz and 7.5--7.8 mHz.
Interestingly enough, we find that the relative importance of the two observed frequencies, as far as the power concentration is concerned,
is inconsistent with the amplitude spectra obtained by Koen et al. (1999) three years before. At the time, the authors did report
small variations in mode amplitude occuring in matter of weeks. Our observations bring evidence for larger variations at longer time-scales
ressembling those reported towards PG\,1605$+$072 by O'Toole et al. (2002).

\subsection{PG\,1605$+$072}
Among sdB stellar pulsators this star seems to be peculiar (Kilkenny et al., 1999): it shows the richest power spectrum (up to 50 modes),
the longest period (550 s), and the lowest surface gravity ($\log$ g $\approx$ 5.25). It has been proposed that this star has
just left the EHB and is now slowly turning into a low mass WD. We obtain the two main frequencies of 2.1 mHz and
2.7 mHz from non-linear sinusoidal fit (see Fig. 2). These large amplitude oscillations are clearly visible in the derived FUV
light curve (see section 5.3) and have the same amplitude than the ones observed in the optical domain (0.25 mag) as
expected for low-g dwarf stars (Kepler et al., 2000).

\subsection{PG\,1613$+$426}
This star is believed to sit near the hot end of the sdB instability trip and is thus expected to show pulsations.
Photometric observations of PG\,1613$+$426 were carried out with the 91--cm Cassegrain telescope of the M. G. {\it Fracastoro}
stellar station of the Catania Astrophysical Observatory in between June and August 2002 collecting more than
45,000 seconds of exposure. These observations allowed to detect a dominant frequency at 6.936 $\pm$ 0.003 mHz
while no additional periods could be detected (Bonanno et al., 2003).

In a single 12,070 s {\it FUSE} observation a non--linear least square fit of a multi--periodic sinusoid allows to
detect the main frequency peak at 7.048 $\pm$ 0.3 mHz with a confidence level above 90 \% (see Fig. 3). Unfortunately
severe aliasing prevented us to identify other possible pulsation modes but this limitation could be avoided
in the future with a different planning of the {\it FUSE} observations.

\section{Time resolved FUV spectroscopy of PG\,1605$+$072}

\subsection{Previous observations}
Several attempts have been made recently to monitor the spectral variations of sdB pulsators, in particular PG\,1605$+$072, as
well as to model them. Although PG\,1605$+$072 is considered the best target from the observational point of view it is also
famous for being a challenging pulsator in terms of interpretation. 

Soon after its discovery (Kilkenny et al., 1999), PG\,1605$+$072 was selected for a detailed quantitative spectral analysis.
The richness of its pulsation spectrum was simultaneously interpreted by Kawaler et al. (1999) as an effect of g--trapped mode in an extremely
fast rotating star: v$_{\rm rot}$ $\approx$ 130 km.s$^{-1}$. This prediction found a remarkable confirmation in the first observation of its optical spectrum
by Heber et al. (1999) who found a line broadening of the order of 39 km.s$^{-1}$. But these data were obtained during an integration of 600 seconds
on target (longer than the main 480 seconds period) and no measurements of radial velocity variations were available to disentangle between
temperature, rotation and Doppler shift.

The possibility of detecting the radial velocity variations
towards this star has been first demonstrated by O'Toole et al. (2000, 2002) using a 2--meter class telescope. Their analysis
was based on the wavelength displacement of the Balmer lines and led them to detect seven oscillation velocities, most corresponding
to previously reported photometric modes. Interestingly enough, adding the velocity amplitudes of the four dominant modes
(around 2.1 mHz and 2.7 mHz) lead then to an observable combined velocity amplitude of roughly 24 km.s$^{-1}$. However, despite the
extended time-series obtained for this observation, a severe limitation on the precise amplitude measurement was set by the very low
resolution available at the time (6 \AA~ at 4,000 \AA). Additional observations with a 4--meter class telescope (Anglo--Australian telescope)
showed an apparent change in pulsation amplitude but, because of the short observing run, any attempt to use these observations for
further asteroseismological work was compromised (Jeffery et al., 2003). To this day, it has been impossible to revisit the Kawaler's prediction
on the rapid rotation of PG\,1605$+$072.

Continuing this pioneering work, O'Toole et al. (2003) presented the detection of line--index variations of Balmer lines.
Because line--index variations are related to both $\delta$ $T$ and $\delta$ $\log$ g, it was argued that such measurement
may be used as an alternative to time--resolved spectroscopic data. Then, using a basic model of stellar pulsation,
they proposed a temperature variation of about 490 K, best consistent with the previous observations of velocity amplitude.

In the following sections, after a brief presentation of the FUV spectrum and interstellar lines observed along
the line of sight, we compute a composite lpv diagram of the star and compare our new results with the literature.

\subsection{The FUV spectrum}
PG\,1605$+$072 is by far the brightest sdB star observed with {\it FUSE} and is thus the most amenable to spectroscopic study.
An interstellar component is clearly identified along the line of sight within which small amounts of molecular hydrogen
are detected: $\log$ N$_{\rm H2}$ $\approx$ 18.7 cm$^{-2}$ with a total b--value of 6.6 km.s$^{-1}$. This diffuse molecular component is observed
through the first 4 rotational levels of the molecular hydrogen and is well described by a single excitation temperature
of 130 K. With such a low excitation temperature and an average velocity shift of $+$23 km.s$^{-1}$ relative to the photospheric lines, it
is unlikely that this diffuse molecular component be physically linked to PG\,1605$+$072.
These numerous H$_{2}$ absorption lines that plague the FUV spectrum make difficult the analysis of the stellar lines but, after
a careful inspection, we can finally identify a handful of useful photospheric lines: N III doublet
at 100.60 nm, S III* at 101.56 nm and 101.58 nm, S IV at 106.27 nm, and S IV* at 107.30 nm and 107.35 nm (see Table 2).
Because the photospheric lines are highly structured and mildly saturated it is not actually possible to derive accurate
estimates of the column density, temperature and b-value for each ion. Another important information can however be gained
from the analysis of the line profile variations diagram (hereafter lpv diagram).

\subsection{The composite lpv diagram}
To derive the lpv diagram towards PG\,1605$+$072, the 4 strongest photospheric lines (Table 2.) are simply co--added
and the alignment is performed by cross-correlating over the numerous H$_{2}$ absorption lines present around each stellar line. 
To optimize the signal--to--noise, we divide the main pulsation period into 8 phase intervals of 60 seconds (T = 480 s; f = 2.1 mHz).
Fig 4. shows the resulting composite lpv diagram.

A remarkable feature of the derived lpv relies in its relative simplicity at variance with the complexity usually
exhibited  by relatively high order (l $>$ 2) non radial modes stars (Mantegazza et al., 2002). Although fine structures might be undetected
at the {\it FUSE} resolution (15 km.s$^{-1}$) the actual lpv is not inconsistent with the presence of a dominant large amplitude
radial mode. To allow for a more quantitative approach  a simultaneous profile fitting of the high ions was performed
for each phase interval assuming that the photospheric lines are well approximated by a simple Voigt profile. The
profile fittings showed that the central velocity of the stellar lines roughly follows a single sinusoid with 17 km.s$^{-1}$ amplitude
which, at face value, translates into a radial amplitude of about 2,600 km (1\% of the stellar radius). We note that this result is an
oversimplification since at least 4 nearby dominant velocity modes are present but it is consistent with velocities
variations reported by O'Toole et al. (2002) using 2m class telescopes and by Woolf et al. (2002) using 4m class telescopes.

Using simultaneous information of photometric and velocity amplitudes, we are also able to derive in a straightforward manner
the phase difference of the maximum flux amplitude relative to maximum radius amplitude and find from a sinusoid fits (Fig. 5)
$\phi$ = 180 $\pm$ 5 degrees. Noteworthy is the fact that this computation
is remarkably consistent with the expected phase opposition which is the signature of an adiabatic process.

In a previous spectroscopic investigation of this star Heber et al. (1999) reported a total b--value of the order of 39 km.s$^{-1}$ but, as
Fig. 4 shows, those studies performed at low speed are biased by the rapid Doppler shift (in a 600 s integration on target, we shall
expect to find a total b--value of the order of two times 17 km.s$^{-1}$ leading to 34 km.s$^{-1}$ close to the reported value of 39 km.s$^{-1}$).
Indeed our systematic profile fitting of the photospheric lines (see phases 2 and 3 in Fig. 4) leads to a much smaller b-value
of 21 $\pm$ 9 km.s$^{-1}$ (2--$\sigma$ error bars). 

From the theoretical point of view, the low value we derive for the total b-value
does not fit easily into the current pulsation models for PG 1605+072 (Kawaler et al., 1999; Kilkenny et al., 1999).
In this model, low--order non--radial pulsations are the signature of trapped modes associated with an equatorial
velocity of 130 km.s$^{-1}$. Such a large velocity is inconsistent with our derived b-value unless we assume a rather special
geometry with the star being observed almost pole--on (i $\approx$ 10 deg). {\it De Facto}, we note also that such a large rotational
velocity would be unique among degenerated dwarf stars. An alternative model has recently been put
forward based on the tentative detection of closely spaced frequencies within the 2.7 mHz peak (Woolf et al., 2002) 
possibly due to rotational
splitting of a multiplet. Then assuming the frequency splitting between modes with successive m values is $\delta\nu$ $\approx$
$\nu_{\rm rot}$/[$l$($l$+1)] $\approx$ 0.01 mHz, where $\nu_{\rm rot}$ is the rotational frequency, and assuming R = 0.28 R$_{\odot}$
(Woolf et al., 2002), we note that a l=1 mode split into m=0, $\pm$1 leads to
V$_{\rm rot}$ = 21--23 km.s$^{-1}$. Indeed, such a rotational velocity seems roughly consistent with our observation but
better quality data are still mandatory for an unambiguous mode identification.

\section{Future prospects for space--based asteroseismology with {\it FUSE}}

The new results reported in this work have been made possible thanks to the unexploited photometric capabilities of the
{\it FUSE} satellite. {\it FUSE} is equipped with two
microchannel plate detectors (90,000 by 200 resolution elements each) and can handle a total counting rate of
the order of  32,000 events per second with a quantum efficiency of 15-30 \% in the 90-120 nm range (Sahnow et al., 2000).
In time-tagged mode operation it is for instance possible to achieve a photometric signal--to--noise ratio of about 1,000
in 30 seconds.

Noteworthy is the fact that {\it FUSE} is not the only instrument that can be diverted to provide photometric data and a few prior
attempts to detect pulsations from space have been reported with {\it HST} towards white dwarf stars
as early as 1995 (see Robinson et al., 1995). Alas, given the present status of servicing mission to {\it HST}, {\it FUSE} will soon be
a unique instrument of its kind. In the coming years, more datasets will become available to the asteroseismological community thanks
to dedicated missions such as {\it COROT} (Baglin et al., 2001) and {\it MOST} (Walker et al., 2003), but will these missions totally
eclipse the potential of the {\it FUSE} satellite for asteroseismology or will {\it FUSE} enter the run?

Table 3. shows a simulation of performances between {\it FUSE}, {\it MOST} and {\it COROT} for
three typical targets: a 15$^{\rm th}$ magnitude white dwarf ($T$ = 40,000 K), a 9$^{\rm th}$ magnitude O9 star, and 6$^{\rm th}$ magnitude
solar type star (G2). The comparison is set on the basis of routine operations: from  a 9,000 seconds continuous exposure with {\it FUSE}
to a 6 months observation with {\it COROT}. 

The reported 2--$\sigma$ detection in the Fourier space has been extrapolated
from prelaunch performances for {\it COROT} (Baglin et al., 2001) and the {\it MOST} in--flight performances we adopt are extrapolated from the recent
photometric results achieved towards Procyon (Matthews et al., 2004) which seem rather modest compared to prelaunch estimates; hence, a
2--$\sigma$ detection of the order of 6 ppm (part per million) is obtained above 2 mHz for a V=0.38 star (Procyon).

To compute the {\it FUSE} detection capabilities we estimated the total fluxes for each target following the reliable
estimates available on--line at the {\it FUSE} web site. Then adding a white noise consistent with the expected signal--to--noise
we buried a 10 mHz mode of constant amplitude in the FUV time--series. In order to obtain realistic time--series we divided each
{\it FUSE} observation into chunks of 9,000 seconds each separated by gaps of 9,000 seconds. We then report the detectability
of the buried mode in the Fourier space.

In conclusion, as Table 3. shows, the estimated performances of the {\it COROT} photometer are several orders of magnitude beyond any
other instrument for stars brighter than the 9$^{\rm th}$ magnitude. But as Table 3. also shows {\it FUSE} is indeed better suited
than any other instrument regarding faint stars studies: for the years to come {\it FUSE} will be the only instrument making faint
(Mv $>$ 13), short period (less than 60 seconds) pulsators amenable to study.
We futher note that its design and mission planning come with three other unique advantages: 1) it can routinely perform several snapshot
observations of 4,000 seconds in a row with a much higher sensitivity than any ground-based facility, 2) it offers the fastest sampling
rate ever achieved in space (up to 1 Hz), and 3) it can provide high--speed time--resolved spectroscopic data in the FUV domain where
many photospheric lines occur. To our knowledge these latter performances will have no equivalent among space telescopes for the next decade.

As of today, about 1,500 different stars have been observed with the {\it FUSE} satellite in time--tagged mode among which dozens of sdB stars
(see http://achive.stsci.edu/fuse/). Each of these dataset can be converted into FUV photometric time--series following the calibration method
described in this paper. Many of these datasets may possibly provide high--speed time--resolved spectroscopic data as well. Noteworthy is the fact
that, given the large spectrum of targeted stars available in the archive, the asteroseismological potential of the {\it FUSE} satellite may
well extend far beyond sdB investigations.

\section{Summary}
   \begin{enumerate}
      \item Important progress has been made possible thanks to the many FUV spectra of dwarf stars
	obtained by {\it FUSE} (Fontaine \& Chayer, 2004). In this paper, we present an original
	way of handling {\it FUSE} data which allows to fully exploit its high--speed photometric capabilities.
	We report on the detection of two dominant modes towards PG\,1219$+$534 (at 6.9 mHz and 7.8 mHz) and
	towards PG\,1605$+$072 (at 2.1 mHz and 2.7 mHz) and we confirm the existence of a single dominant
	mode towards PG\,1613$+$426 at 7.048 $\pm$ 0.3 mHz.
      \item We further report on the Doppler shift of the photospheric lines at the surface
	of PG\,1605$+$072 with 17 km.s$^{-1}$ amplitude. The time--resolved spectroscopy
	of this star allows us to derive the projected broadening of the photospheric lines
	with unprecedented accuracy: v.sini = 21 $\pm$ 9 km.s$^{-1}$ (2--$\sigma$ error bar). 
	We note that this result is only marginally consistent with the current theoretical expectations that 
	the rapid rotation of PG\,1605$+$072 is the source of its pulsation pattern. Thanks to these new data
	we are also able for the very first time to observe the phase opposition between the maximum radius and the
	maximum velocity as expected if adiabatic pulsations are taking place in the envelope of PG\,1605$+$072.
      \item Finally, we demonstrate that {\it FUSE} is particularly well suited for asteroseismological investigation
	of dwarf stars in general. Comparing the actual {\it FUSE} photometric performances with {\it MOST} and {\it COROT}, we
	conclude that {\it FUSE} is a unique instrument for the study of faint ($Mv$ $>$ 10), hot (T $>$ 20,000 K)
	pulsating objects and should be considered an alternative to the {\it COROT} mission in that observation domain.
   \end{enumerate}

\begin{acknowledgements}
      We are very grateful to James Caplinger and Bryce Roberts from the {\it FUSE} staff for valuable discussions on the {\it FUSE}
detectors and mission planning. Kuassivi is especially grateful to Farid Abdelfettah, Sylvie Andr\'e,
Jean--michel Arslanian, and Olivier Rulli\`ere from the AZimov association for enlightening comments
about the general science reported in this paper. The work at the Institut d'Astrophysique de Paris
was supported by CNES/CNRS/UPMC. The work at the Osservatorio Astrofisico di Catania was supported by the
Italian Ministero dell'Istruzione, Universit\`a e Ricerca.
\end{acknowledgements}

\newpage
%------------------------------------------------------------

%       TABLE 1

   \begin{table*}
      \caption[]{Observation summary.}
         \label{Tab1}
     $$ 
         \begin{array}{p{0.2\linewidth}rrllccc}
            \hline
            \noalign{\smallskip}
                    & \textrm{R.A.} & \textrm{Decl.} & \textrm{Mag} &\textrm{Date of}          &\textrm{Exposure}   & \textrm{Program}        \\
            Stars   &\textrm{J2000} & \textrm{J2000} & \textrm{V}   &\textrm{observation$^{*}$}&\textrm{time}       & \textrm{ID}\\
            \noalign{\smallskip}
            \hline
            \noalign{\smallskip}
            PG\,1219$+$534 &12~21~29    & $+$53~04~37   & 12.4 & \textrm{2001--01--16} & 6,300  &\textrm{B033}\\
            PG\,1605$+$072  &16~08~04    & $+$07~04~29   & 12.8 & \textrm{2001--07--27} & 4,700  &\textrm{B033}\\
            PG\,1613$+$426 &16~14~47    & $+$42~27~36   & 14.4 & \textrm{2002--07--15} & 12,070 &\textrm{Z904}\\
            \hline
         \end{array}
     $$ 
\begin{list}{}{}
\item[$^{\mathrm{*}}$] All the targets have been observed through the large {\it FUSE} aperture (LWRS).
\end{list}
   \end{table*}

%---------------------------------------------------------------
%------------------------------------------------------------

%       TABLE 2

   \begin{table*}
      \caption[]{Identification of some photospheric lines in the spectrum of PG\,1605$+$072.}
         \label{Tab2}
     $$ 
         \begin{array}{p{0.05\linewidth}ccr}
            \hline
            \noalign{\smallskip}
             Species       & \textrm{$\lambda$} & \textrm{f$_\lambda$} & \textrm{Equivalent Width}        \\
                           & \textrm{nm}        &                      & \textrm{m\AA}                   \\
            \noalign{\smallskip}
            \hline
            \noalign{\smallskip}
            N\,{\small III}  	   &  100.60          &  3.76e-2             & \textrm{105 $\pm$ 10}\\
            S\,{\small III$^{*}$}  &  101.56          &  ...                 & \textrm{30 $\pm$ 5}\\
            S\,{\small III$^{*}$}  &  101.58          &  ...                 & \textrm{30 $\pm$ 5}\\
            S\,{\small IV}         &  106.27          &  5.8e-1              & \textrm{45 $\pm$ 6}\\
            S\,{\small IV$^{*}$}   &  107.30          &  5.2e-1              & \textrm{120 $\pm$ 10}\\
            S\,{\small IV$^{*}$}   &  107.35          &  5.7e-2              & \textrm{30 $\pm$ 5}\\
            \hline
         \end{array}
     $$ 
  \end{table*}

%---------------------------------------------------------------
%------------------------------------------------------------

%       TABLE 3

   \begin{table*}
      \caption[]{{\it FUSE}, {\it MOST}, and {\it COROT} photometric performances.}
         \label{Tab3}
     $$ 
         \begin{array}{p{0.2\linewidth}rcccc}
            \hline
            \noalign{\smallskip}
                  		& \textrm{{\it FUSE}} & \textrm{{\it FUSE}} & \textrm{{\it MOST}}  &  \textrm{{\it COROT}}      \\
                           	& \textrm{Archive$^{*}$}    & \textrm{Legacy$^{**}$}     &                &                       \\
            \noalign{\smallskip}
            \hline
            \noalign{\smallskip}
            Exposure time        &  \textrm{9,000 s}      &  \textrm{4 days}        & \textrm{32 days}      &  \textrm{150 days}\\
            Frequency resolution &  \textrm{110 $\mu$Hz}  &  \textrm{2.5 $\mu$Hz}   & \textrm{0.36 $\mu$Hz} &  \textrm{0.08 $\mu$Hz}\\
            Maximum sampling rate&  \textrm{1 Hz}          &  \textrm{1 Hz}           & \textrm{83 mHz}	    &  \textrm{31 mHz}\\
            \hline
            \noalign{\smallskip}
            2--$\sigma$ detection (ppm)\\
	    WD, Mv = 15     &  \textrm{5,000}         &  \textrm{800}       & \textrm{...}      & \textrm{...}   \\
            O9 star, Mv = 9 &  \textrm{700}           &  \textrm{100}       & \textrm{4,600}    & \textrm{4.8}  \\
            G2 star, Mv = 6 &  \textrm{...}           &  \textrm{...}       & \textrm{290}      & \textrm{1.2}  \\
            \hline
         \end{array}
     $$ 
  \begin{list}{}{}
\item[$^{\mathrm{*}}$] An exposure time of 9,000 is typical of many archived {\it FUSE} spectra.
\item[$^{\mathrm{**}}$] An exposure of 4 continuous days assuming a duty cycle of 50 \% leads to 180,000 seconds of
total exposure time. Such observation program has not been achieved yet with {\it FUSE} towards
a pulsating stars but may be planned as part of a {\it FUSE Legacy} program.
\end{list}
  \end{table*}

%---------------------------------------------------------------

%\newpage
%_________________FIGURE 1_____________________
%   \begin{figure*}[htbp]
%   \centering
%   \includegraphics[width=\textwidth]{figure4.ps}
%   \caption{Image of one {\it FUSE} detector while observing PG\,1605$+$072 the LWRS aperture on detectors 1A
%and 1B. The entire spectral range is shown (LIF 1A: 98.7 -- 108.2 nm, LIF 1B: 109.4 -- 118.8 nm, SIC 1A: 100.4 --
%109.1 nm, SIC 1B: 90.5 -- 99.2 nm).Signs and labels are plotted over areas of the detector that cannot be used for
%photometric purposes: {\bf A} labels indicate the strongest H\,{\small I} airglow emission lines at 102.5 nm, {\bf W}
%labels designate aeras of the
%detector affected by the moving worm structure, arrows point to the strongest stellar and interstellar absorption
%lines ({\bf 1}: C\,{\small III} at 108.4 nm, {\bf 2}: H\,{\small I} at 97.2 nm, {\bf 3}: H\,{\small I} at 94.9 nm,
%{\bf 4}: H\,{\small I} at 93.7 nm, {\bf 5}: H\,{\small I} at 93.1 nm).}
%              \label{Fig1}%
%    \end{figure*}
%______________________________________________

\newpage
%_________________FIGURE 1_____________________
   \begin{figure*}[htbp]
   \centering
   \includegraphics{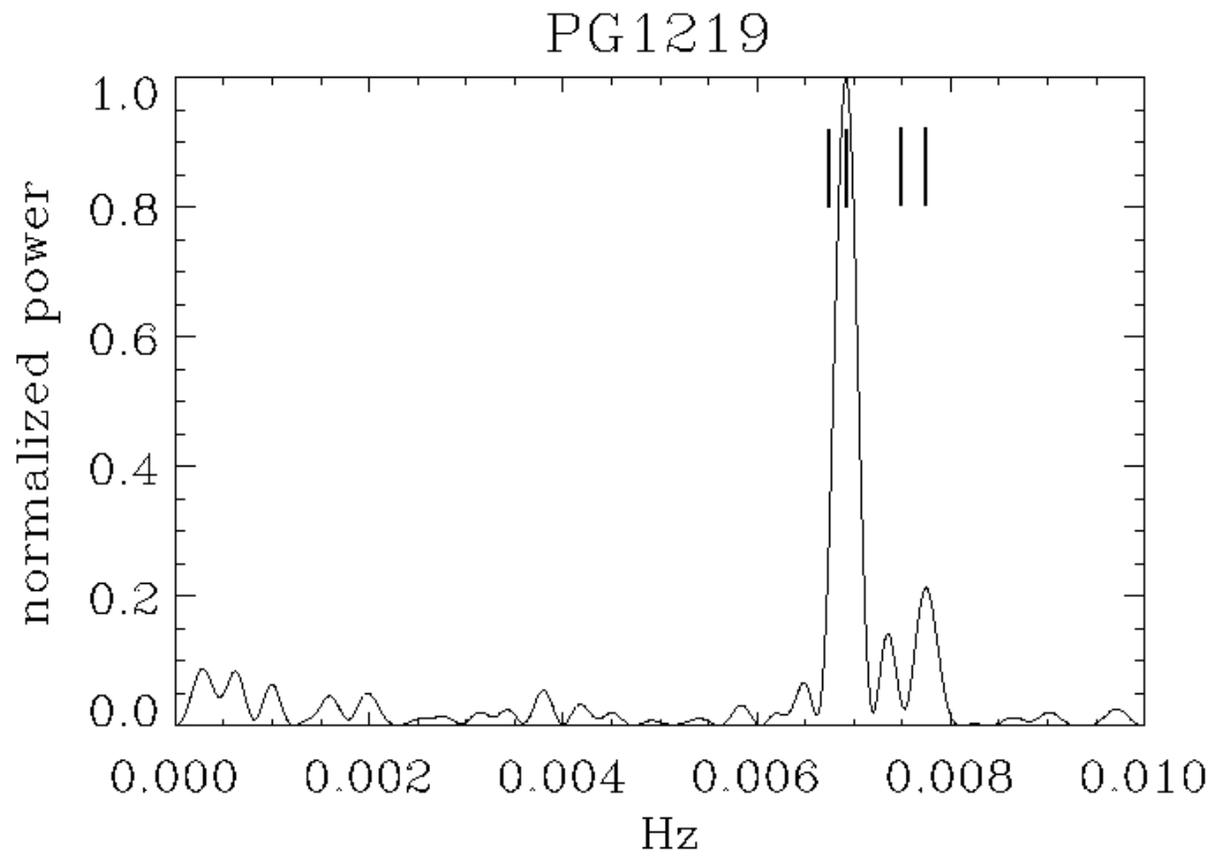}
   \caption{DFT periodogram obtained towards PG\,1219$+$534. The vertical lines show the ground based frequencies (Koen et al. 1999).}
              \label{Fig1}%
    \end{figure*}
%______________________________________________

%_________________FIGURE 2_____________________
   \begin{figure*}[htbp]
   \centering
   \includegraphics{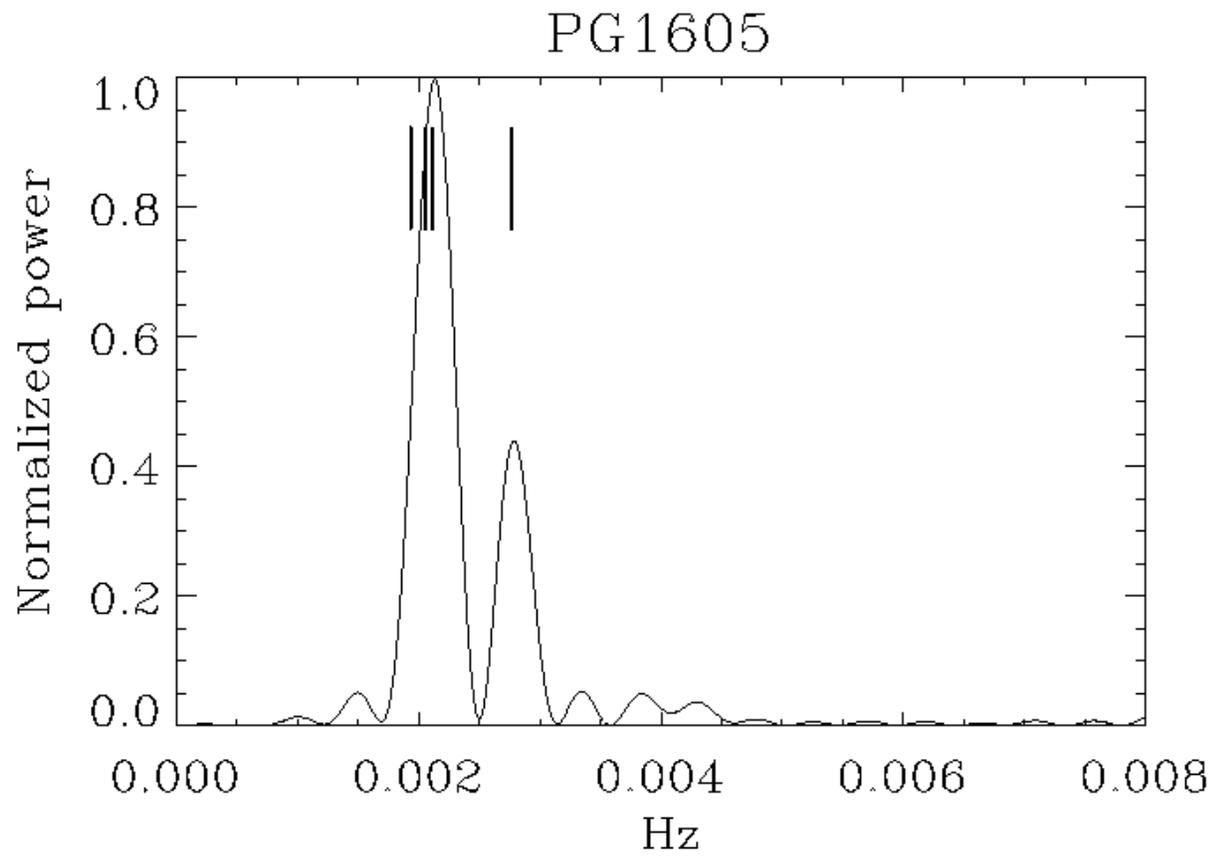}
   \caption{DFT periodogram obtained towards PG\,1605$+$072.  The vertical lines show the four dominant ground based frequencies in this
   frequency domain (O'Toole et al., 2002).}
              \label{Fig2}%
    \end{figure*}
%______________________________________________

\newpage
%_________________FIGURE 3_____________________
   \begin{figure*}[htbp]
   \centering
   \includegraphics{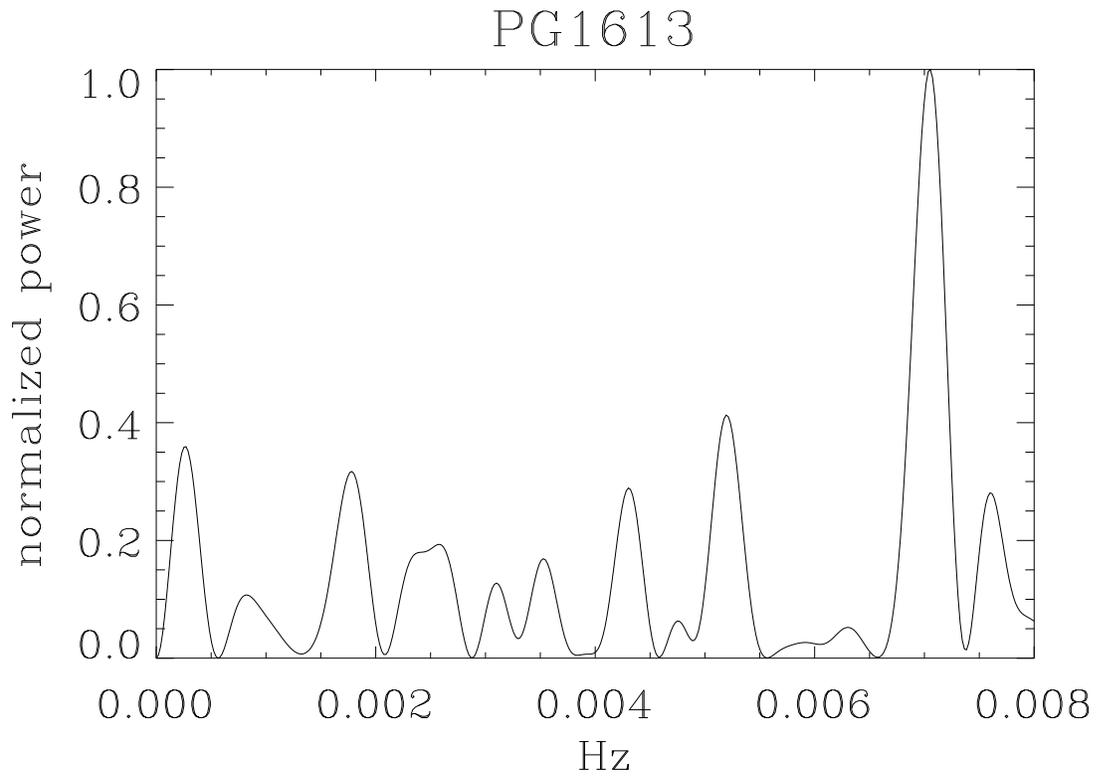}
   \caption{ DFT periodogram obtained towards PG\,1613$+$426. The dominant peak coincides with the previous ground based detection at 6.9 mHz 
   (Bonanno et al.,2003).}
              \label{Fig3}%
    \end{figure*}
%______________________________________________
%\newpage
%_________________FIGURE 3_____________________
%   \begin{figure*}[htbp]
%   \centering
%   \includegraphics{psd1605.ps}
%   \caption{Lomb--Scargle periodogram obtained towards PG\,1613$+$426.}
%              \label{Fig3}%
%    \end{figure*}
%______________________________________________

\newpage
%_________________FIGURE 4_____________________
   \begin{figure*}[htbp]
   \centering
   \includegraphics[width=12cm]{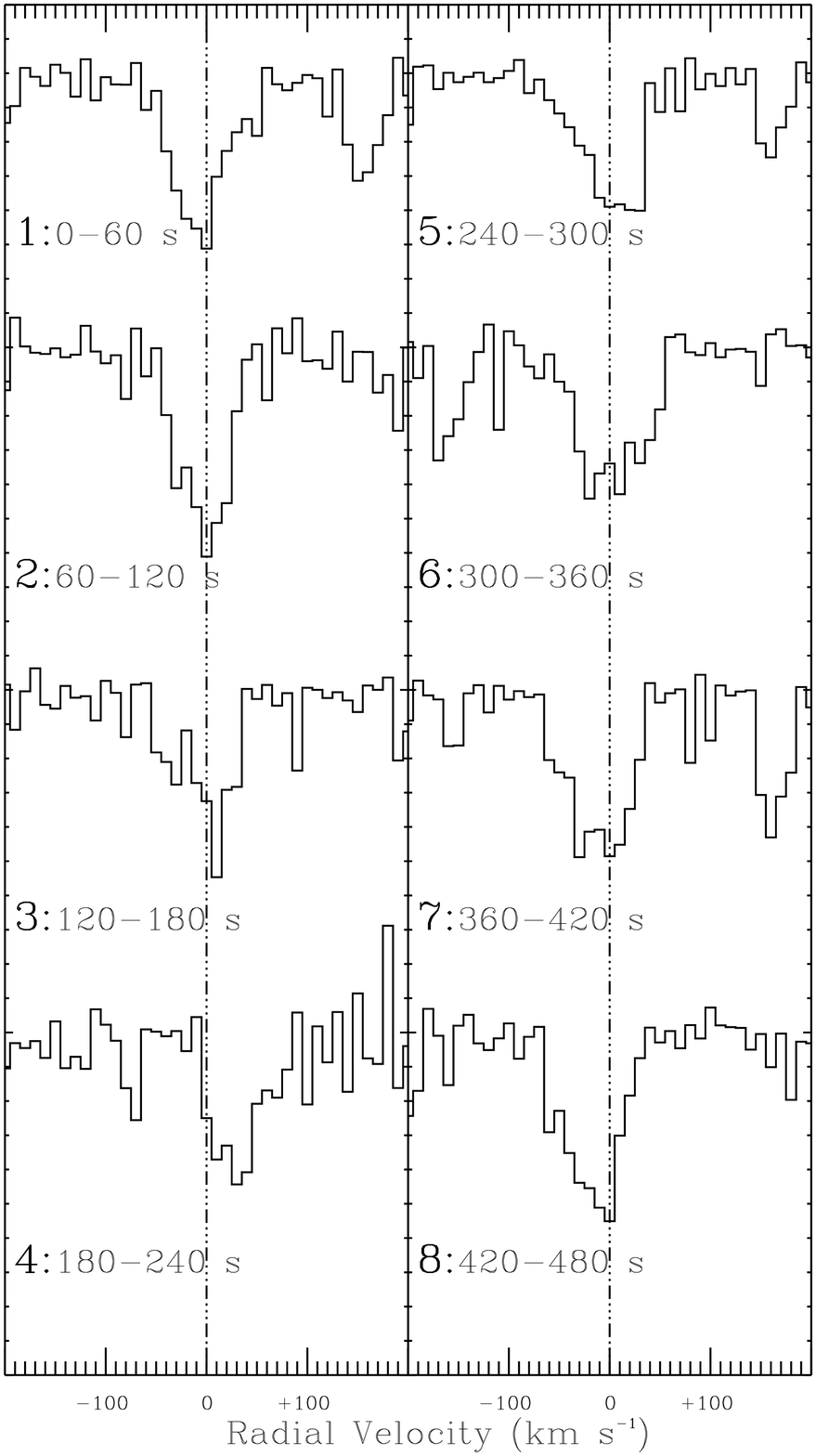}
   \caption{Line profile variations observed at the surface of PG\,1605$+$072. Each panel represents a 60 seconds
integration bin and each plotted profile, binned by 10 km.s$^{-1}$, is an average profile obtained from the coaddition
of 4 strong photospheric lines: N\,{\small II} doublet at 100.60 nm, S\,{\small III*} at 101.56 nm, S\,{\small IV}
at 106.27 nm, and S\,{\small IV*} at 107.30 nm. The central dotted line, around which the profiles move, shows
the average zero velocity of photospheric species.}
              \label{Fig4}%
    \end{figure*}
%______________________________________________

\newpage
%_________________FIGURE 5_____________________
   \begin{figure*}[htbp]
   \centering
   \includegraphics[width=\textwidth]{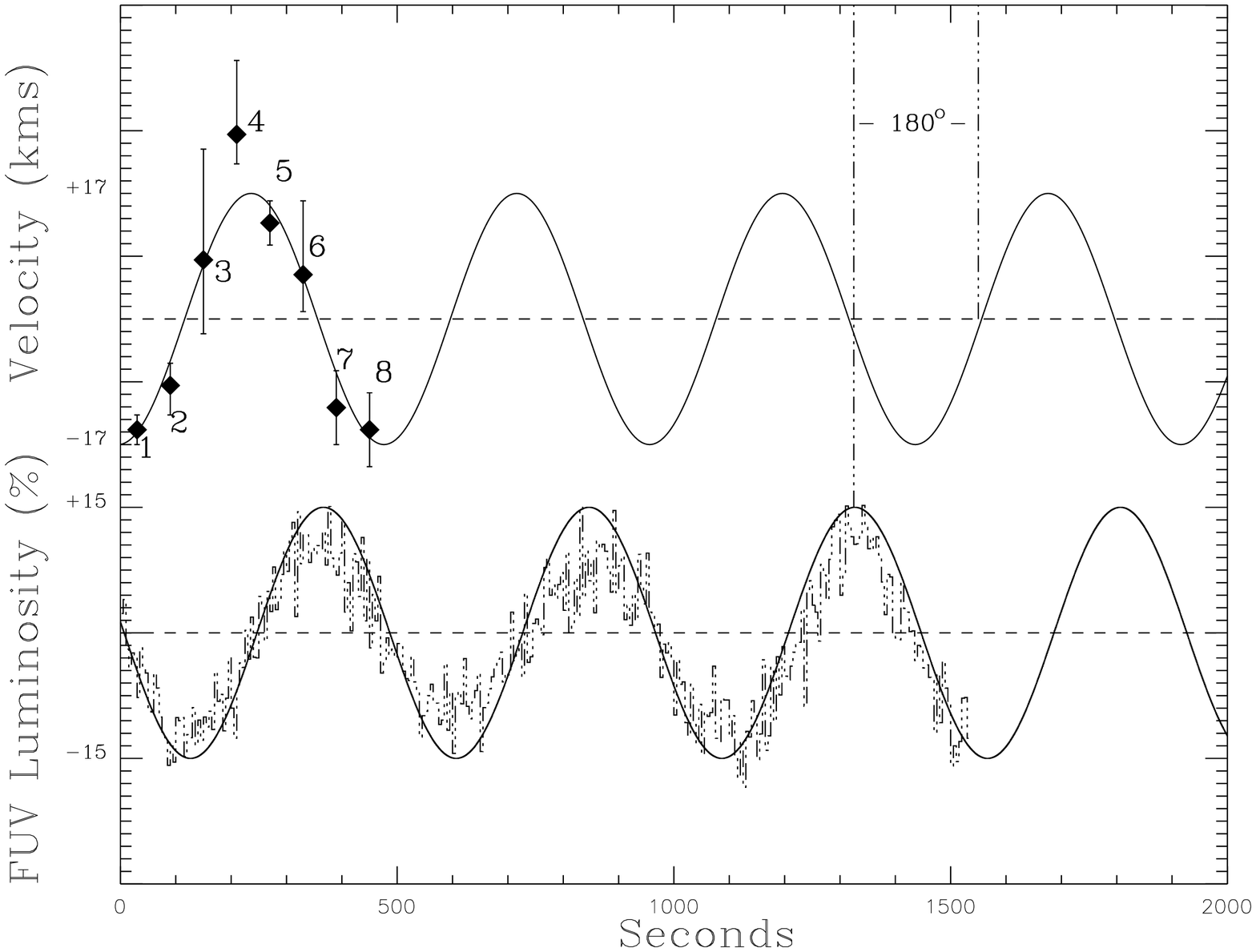}
   \caption{Radial velocity and luminosity variation curves derived from the PG\,1605$+$072 observation. The radial
velocities are derived from profile fitting of the lpv diagram (plotted error bars are 1--$\sigma$) while the FUV photometric data shown
above are directly obtained from our special calibration procedure (see section 3). The two vertical dotted lines
helps to compare the maximum luminosity and minimum radius position in phase. We note that both maxima are in phase
opposition as expected from adiabatic pulsations.}
              \label{Fig5}%
    \end{figure*}
%______________________________________________


\begin{thebibliography}{}

  \bibitem[2001]{baglin} Baglin, A., Auvergne, M., Catala, C., Michel, E., \& {\it COROT} Team 2001,
      in SOHO/GONG Workshop

   \bibitem[2003]{bonanno} Bonanno, A., Catalano, S., Frasca, A., Mignemi, G., \& Patern\`o, L. 2003, A\&A, 398, 283

   \bibitem[2000]{brown} Brown, T.M., Bowers, C.W., Kimble, R.A., Sweigart, A.V., \& Ferguson, H.C., 2000, \apj, 532, 308
      A\&A, 200, 58

   \bibitem[1996]{charpinet} Charpinet, S., Fontaine, G., Brassard, P., \& Dorman, B. 1996, \apj, 471, 103

   \bibitem[2004]{Fontaine} Fontaine, G. \& Chayer, P. 2004, ArXiv Astrophysics e--print, astr--ph/0411091
   \bibitem[1986]{heber86} Heber, U., Kudritzki, R.P., 1986, A\&A, 169, 244

   \bibitem[1999]{heber99} Heber, U., Reid, I.N., \& Werner, K. 1999, A\&A, 348, L25
   
   \bibitem[2004]{Jeffery04} Jeffery, C.~S. 2004, \apss, 291, 403
   
   \bibitem[1999]{kawaler} Kawaler, S.D. 1999, 11th European Workshop on White Dwarfs, ASP conf. Series \# 169, ASP, p158
  
   \bibitem[2000]{kepler} Kepler, S.O., Robinson, E.L., Koester, D., Clemens, J.C., Nather, R.E., \& Jiang, X.J. 2000,
	\apj, 539, 379

   \bibitem[1997]{kilkenny97} Kilkenny, D., Koen, C., O'Donoghue, D., \& Stobie, R.S. 1997, MNRAS, 285, 640

    \bibitem[1999]{kilkenny99} Kilkenny, D., Koen, C., O'Donoghue, D., van Wyk, F., Larson, K.A., Shobbrook, R.,
	Sullivan, D.J., Burleigh, M.R., Dobbie, P.D., \& Kawaler, S.D. 1999, MNRAS, 303, 525

   \bibitem[1999]{koen} Koen, C., O'Donoghue, D., Pollacco, D.L., \& Charpinet, S. 1999, MNRAS, 305, 28

   \bibitem[2004]{Kuassivi} Kuassivi, Bonanno, A., Ferlet, R., Roberts, B., \& Caplinger, J. 2004, ArXiv Astrophysics e--print, astro--ph/0411371

   \bibitem[2002]{mante} Mantegazza, L., \& Poretti, E. 2002, A\&A, 396, 911

   \bibitem[2004]{matthews} Matthews, J.M., Kusching, R., Guenther, D.B., Walker, G.A.H., Moffat, A.F.J., Rucinski, S.M.
	Sasselov, D., \& Weiss, W.W. 2004, Nature, 430

   \bibitem[2000]{moos} Moos, H.W., et al. 2000, \apjl, 538, L1
 
   \bibitem[2000]{Otoole} O'Toole, S.J. et al. 2000, \apj, 537, L53

     \bibitem[2002]{Otoole} O'Toole, S.J., Bedding, T.R, Kjeldsen, H., Dall, T.H. \& Stello, D. 2002, MNRAS,
     334, 471

   \bibitem[2003]{Otoole} O'Toole, S.J., J{\o}rgensen, M.S., Kjeldsen, H., Bedding, T.R, Dall, T.H. \& Heber, U. 2003, MNRAS,
   340, 856

    
   \bibitem[1995]{Robinson} Robinson, E.L., Mailloux, T.M., Zhang, E., Koester, D., Stiening, R.F., Bless, R.C.,
  	Percival, J.W., Taylor, M.J. \& van Citters, G.W., 1995, \apj, 438, 908
   
   \bibitem[2000]{sahnow} Sahnow, D.J., \& the {\it FUSE} team 2000, \apjl, 538, L7

 %  \bibitem[1982]{scargle} Scargle, J.D. 1982, \apjl, L263, 865

   \bibitem[2003]{walker} Walker, G. et al. 2003, PASP, 115, 1023

   \bibitem[2002]{woolf} Woolf, V.M., Jeffery, C.S., \& Pollacco, D.L. 2002, MNRAS, 329, 497

\end{thebibliography}
\end{document}